\documentclass[a4paper,fleqn]{cas-dc}

\usepackage{graphicx}
\usepackage[utf8]{inputenc}
\usepackage{bm}  
\usepackage{siunitx}
\usepackage[svgnames]{xcolor}
\usepackage{tabularx}
\usepackage{xspace}  
\usepackage{amsfonts}
\usepackage{amssymb}
\usepackage{amsmath}
\usepackage{natbib}

\newcommand*{\vb}[1]{\boldsymbol{#1}}  
\newcommand*{\dd}{\mathrm{d}}  

\newcommand*{\laplacian}{\nabla^2}
\newcommand*{\gradient}{\vb{\nabla}}

\newcommand*{\uvec}{\vb{u}}
\newcommand*{\Bvec}{\vb{B}}

\newcommand*{\rvec}{\vb{r}}

\newcommand*{\gvec}{\vb{g}}
\newcommand*{\lconv}{\bar{\ell}_\mathrm{u}}
\newcommand*{\Hs}{\mathcal{H}_s}
\newcommand*{\CMB}{\mathrm{CMB}}
\newcommand*{\Nmax}{N_\mathrm{max}}
\newcommand*{\Ylm}{Y_l^m}

\newcommand*{\fdip}{f_\mathrm{dip}}
\newcommand*{\tdip}{t_\mathrm{dip}}
\newcommand*{\Hcmb}{H_\CMB}

\newcommand*{\dq}{\delta q}
\newcommand*{\Bpol}{B_{\mathrm{pol}}}
\newcommand*{\Btor}{B_{\mathrm{tor}}}
\newcommand*{\Ekin}{E_{\mathrm{kin}}}
\newcommand*{\Emag}{E_{\mathrm{mag}}}

\newcommand*{\mean}[1]{\langle #1 \rangle}

\newcommand*{\Rey}{\mathrm{Re}}
\newcommand*{\Rm}{\mathrm{Rm}}
\newcommand*{\Ek}{\mathrm{E}}
\newcommand*{\Ra}{\mathrm{Ra}}
\newcommand*{\Rac}{\mathrm{Ra}_c}
\newcommand*{\Ro}{\mathrm{Ro}}
\newcommand*{\Rol}{\mathrm{Ro}_\ell}
\newcommand*{\Prr}{\mathrm{Pr}}
\newcommand*{\Pm}{\mathrm{Pm}}

\begin{document}

\shorttitle{Magnetic reversals in a geodynamo model with a stably-stratified layer}
\shortauthors{M\"uller et al.}

\title[mode=title]{Magnetic reversals in a geodynamo model with a stably-stratified layer}

\author[1]{Nicol\'as P. M\"uller}
\cormark[1]
\ead{nicolas.muller@lpp.polytechnique.fr}
\author[1,2]{Christophe Gissinger}
\author[1]{François P\'etr\'elis}
\affiliation[1]{organization={Laboratoire de Physique de l'\'Ecole normale sup\'erieure, ENS, Universit\'e PSL, CNRS, Sorbonne Universit\'e, Universit\'e Paris Cit\'e}, postcode={F-75005}, city={Paris}, country={France}}
\affiliation[2]{organization={Institute Universitaire de France (IUF)}, city={Paris}, country={France}}


\maketitle

\begin{abstract}
  We study the process of magnetic reversals in the presence of a stably-stratified layer below the core-mantle boundary using direct numerical simulations of the incompressible magnetohydrodynamics equations under the Boussinesq approximation in a spherical shell. We show that the dipolar-multipolar transition shifts to larger Rayleigh numbers in the presence of a stably-stratified layer, and that the dipolar strength of the magnetic field at the core-mantle boundary increases due to the skin effect. By imposing an heterogeneous heat flux at the outer boundary, we break the equatorial symmetry of the flow, and show that different heat flux patterns can trigger different dynamo solutions, such as hemispheric dynamos and polarity reversals. 
  Using kinematic dynamo simulations, we show that the stably-stratified layer leads to similar growth rates of the dipole and quadrupole components of the magnetic field, playing the role of a conducting boundary layer, favouring magnetic reversals, and a dynamics predicted by low-dimensional models.
\end{abstract}

\section{Introduction}
\label{sec:introduction}

Planetary magnetic fields are generated and maintained through the dynamo process, mechanism through which kinetic energy is converted into magnetic energy, being thus able to sustain a large scale magnetic field \citep{Moffatt1978,Landeau2022,Jones2011}. This mechanism takes place in the core of planets, where the turbulent motion of an electrically conducting fluid is mostly driven by convection and strongly affected by rotation. In the solar system, most planets produce a self-sustained magnetic field, such as the Earth, Mercury and the gaseous giants. Due to differences in the internal composition and size of these planets, the properties of their magnetic fields vary, including their intensity, tilt angle relative to the rotation axis, degree of axial symmetry, and north-south symmetry, among others.
At present, the Earth has a magnetic field with a strong dipolar component with tilting angle of about $11^{\circ}$ and an intensity at its surface that varies between $0.25$ and $0.65$ G. 

One of the most interesting phenomena of the geomagnetic field is its ability to reverse polarity in an irregular and chaotic manner \citep{Lowrie2013}.
Over the past $80$ million years, Earth's magnetic field has reversed on average with a frequency of approximately $4$ Myr$^{-1}$, with individual reversal events typically lasting around $20,000$ years. 
However, both the frequency and duration of reversals vary significantly: some intervals, known as superchrons, can span tens of millions of years without any reversals, while others show more frequent and irregular polarity changes. 
Understanding this process has been one of the main objectives of direct numerical simulations (DNS) of the geodynamo model \citep{Braginsky1995,Glatzmaier1995,Christensen2001,Kutzner2002,Li2002,Wicht2004,Takahashi2005,Driscoll2009,Olson2009,Amit2010,Olson2011,Olson2014,Sheyko2016,Menu2020}.
Parametric studies on this problem suggested that the transition from stable dynamos dominated by a strong dipolar component to reversing solutions is controlled by the relative amplitude of inertia and Coriolis force \citep{Kutzner2002,Christensen2006}. As the flow becomes more turbulent, the reversal frequency increases. 
More precisely, this transition has been proposed to be controlled by the local Rossby number defined as $\Rol = \Ro \lconv / \pi$, with $\Ro = U/(L\Omega)$ the Rossby number, with $U$ the characteristic velocity, $L$ the characteristic length scale, $\Omega$ the rotation rate, and $\lconv$ the characteristic spherical harmonic degree of the flow. Numerical simulations first suggested that the transition takes place around a critical value $\Rol \approx 0.1$, with smaller values of $\Rol$ corresponding to stable dipolar solutions. The reversals found for values $\Rol>0.1$ are typically non-dipolar, meaning that the dipole component is relatively small with respect to the quadrupole and higher order modes, which make these kind of reversals not Earth-like. This regime is usually known as multipolar dynamo. 
There has been since then an interest to find a regime of parameters in which Earth-like magnetic reversals are observed. In particular, simulations show that Earth-like magnetic field reversals occur only within a very narrow range of local Rossby number around $0.1$, a degree of fine-tuning that seems unlikely to have been maintained over hundreds of millions of years in Earth's core. 
Furthermore, it was later observed that the local Rossby number fails to describe the dipolar-multipolar transition for some range of parameters, and other control parameters have been proposed, such as the ratio between kinetic and magnetic energies \citep{Menu2020,Zaire2022,Frasson2025}. In particular, \citet{Tassin2021} showed that reversing dynamos occur when this ratio is larger than $\mathcal{O}(1)$, which is much larger than the typical values estimated for the Earth's core of $10^{-3}$. 
In the recent work of \citet{Jones2025}, it was shown that it is possible to obtain Earth-like reversals while maintaining realistic kinetic-to-magnetic energy ratios by increasing the magnetic Prandtl number together with a corresponding increase in the thermal Prandtl number, preserving realistic transport levels.

The exact nature of convective motion in Earth's outer core remains uncertain. Standard models typically assume that convection occurs throughout the entire region between the inner-core boundary (ICB) and the core-mantle boundary (CMB).
However, recent seismic and geomagnetic measurements have questioned this assumption, suggesting the existence of a stably-stratified layer (SSL) below the CMB \citep{Helffrich2010,Helffrich2013,Buffett2016,Kaneshima2018}. The origin of this layer may be either thermal or compositional in nature, potentially resulting from the accumulation of light elements below the CMB or the high thermal conductivity of the liquid metal relative to the heat flux at the CMB, leading to a sub-adiabatic temperature gradient \citep{Pozzo2012}. 
The existence of such layer in Earth's core remains a subject of debate, with its estimated size ranging from $\Hs=0$ km to $\Hs=450$ km, and stratification strengths ranging from $N = 0$ to $N = 50 \Omega$, where $N$ is the buoyancy frequency \citep{Christensen2018,Gastine2020}. The presence of this layer has also been predicted in other planetary systems, such as Mercury, Saturn or Jupiter, often used to explain different large-scale structures of the magnetic field \citep{Christensen2006a,Moore2018,Takahashi2019,Yan2021,Guervilly2022}. 
One of the main effects of a SSL below the CMB is that high-order modes of the magnetic field are smoothed out due to the skin effect, leading to a magnetic field dominated by low-order modes.  
The recent work of \citet{Aubert2025} proposed a mechanism that differs from the classical approaches based on the levels of inertia of the flow. The proposed kinematic mechanism suggests that in the presence of a stable top layer, there is a competition between the subsurface upwelling (generated in the convective region) and the surface circulation (generated in the stable top layer).

In addition to the possible presence of a SSL below the CMB, seismic measurements and laboratory experiments suggest that the heat flux at the CMB is not uniform, but it rather exhibits a complex spatial distribution \citep{Su1994,Sumita2002}. 
DNS of the geodynamo problem have shown that an heterogeneous heat flux at the CMB can modify the morphology of the magnetic field, trigger polarity reversals, and control their frequency \citep{Glatzmaier1999,Olson2010,Gissinger2012,Christensen2018,Yan2023,Terra-Nova2024,Frasson2025}. 
In particular, the non-axisymmetric structure of the geomagnetic field at the CMB are often associated to lateral heterogeneous heat flux. 

In this work, we study the influence of a stably-stratified layer below the CMB on the geodynamo process, and its effects on polarity reversals. We characterise the dipolar-multipolar transition for different stably-stratified layers, and explore the effects of an axisymmetric heterogeneous heat flux at the CMB that breaks the equatorial symmetry of the flow. 
The goal is to understand how breaking equatorial symmetry influences the large-scale magnetic field and promotes the emergence of Earth-like polarity reversals. We finally use a low-dimensional model and kinematic dynamo simulations to provide a better interpretation of our results. 

\section{Model}
\label{sec:model}

\begin{figure}
  \centering
  \includegraphics[width=\linewidth]{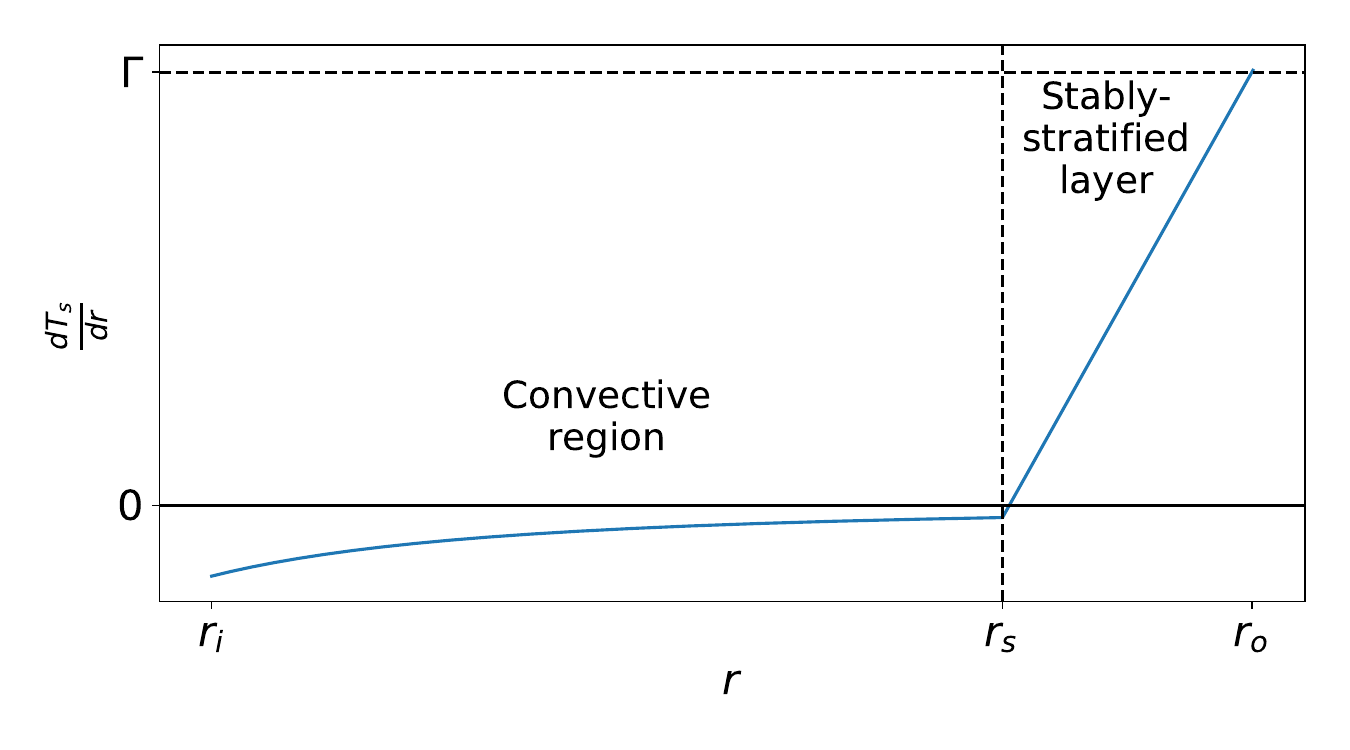}
  \caption[]{ 
  Schematic picture of the static temperature gradient used to reproduce a mixed system, with a convective region and a stably-stratified layer described by \eqref{eq:static_temperature_gradient}. 
    }
    \label{fig:temperature_profile}
\end{figure}

We model the dynamics of an electrically conducting fluid using the incompressible magnetohydrodynamics (MHD) equations under the Boussinesq approximation in a spherical shell of inner radius $r_i$ and outer radius $r_o$. The aspect ratio is defined as $\chi = r_i/r_o$ and is fixed to $0.35$ as the case of Earth. We express the length in units of the gap $L = r_o - r_i$, and the time in units of $L^2/\nu$, with $\nu$ the kinematic viscosity. The magnetic field $\Bvec$ is expressed in units of $\sqrt{\rho \mu \eta \Omega}$, with $\rho$ the mean fluid density, $\mu$ the magnetic permeability, $\eta$ the magnetic diffusivity and $\Omega$ the rotation rate. The velocity field $\uvec$ is expressed in units of $\nu/L$ and the temperature $T$ in units of $q_o L / k$ with $q_o$ the mean heat flux at the CMB and $k$ the thermal conductivity. 
These units are used to obtained the dimensionless MHD equations \citep{Braginsky1995}
\begin{align}
    \partial_t \uvec + \uvec \cdot \gradient \uvec 
    &= - \gradient p + \laplacian \uvec - \frac{2}{\Ek} \hat{z} \times \uvec \notag \\
    &\quad + \frac{\Ra }{\Ek} \frac{r}{r_o} \theta \hat{r} + \frac{1}{\Ek \Pm} (\gradient \times \Bvec) \times \Bvec, \label{eq:ns} 
\end{align}
\vspace{-\baselineskip}
\begin{gather}
    \partial_t \Bvec 
    = \gradient \times (\uvec \times \Bvec) + \frac{1}{\Pm} \laplacian \Bvec, \label{eq:induction} \\
    \partial_t \theta + \uvec \cdot \gradient \theta 
    = - u_r \frac{\dd T_s}{\dd r} + \frac{1}{\Prr} \laplacian \theta, \label{eq:heat} \\
    \gradient \cdot \uvec = \gradient \cdot \Bvec = 0, \label{eq:incompressible}
\end{gather}
where $p$ the total pressure, and $T = T_s + \theta$ the total temperature with $T_s$ the static temperature profile and $\theta$ the temperature fluctuations.
We define the dimensionless numbers using the standard convention: the Ekman number $\Ek = \nu / (\Omega L^2)$, the thermal Prandtl number $\Prr = \nu/\kappa$, the magnetic Prandtl number $\Pm = \nu/\eta$
and the Rayleigh number $\Ra = \alpha g_o q_o L^2 / (\nu k \Omega)$ with $q_o$ the mean heat flux at the CMB, $k$ the thermal conductivity, $\alpha$ the thermal expansion coefficient, 
and the gravity acceleration defined as $\gvec = g_o r/r_o \hat{r}$ with $g_o$ the value at the outer core. 
We use no-slip boundary conditions for the velocity field at both the ICB and CMB, constant temperature at the ICB, fixed heat flux $q_o$ at the CMB, a conducting inner core, and an insulating mantle.

\begin{figure}
  \centering
    \includegraphics[width=\linewidth]{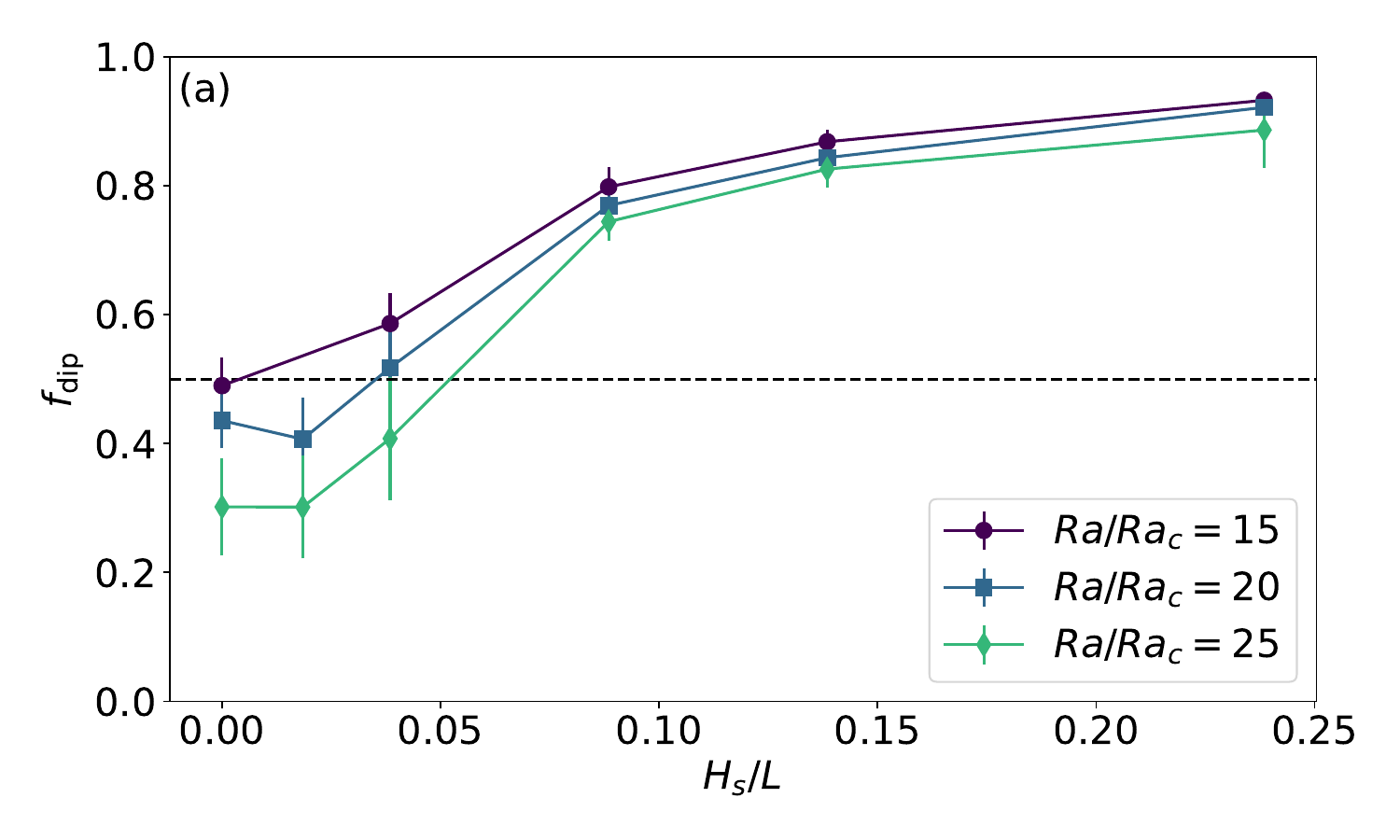}
    \includegraphics[width=.8\linewidth]{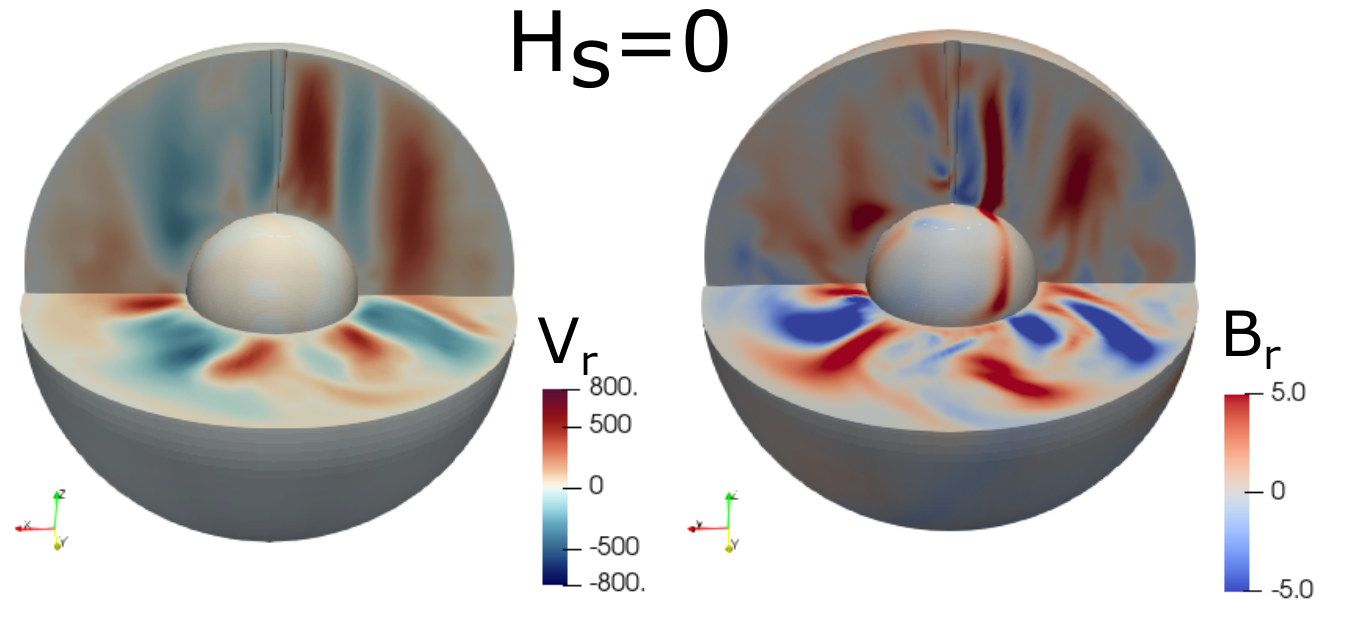}
    \begin{picture}(0,0)
      \put(-200,80){\small (b)} 
    \end{picture}
    \includegraphics[width=.8\linewidth]{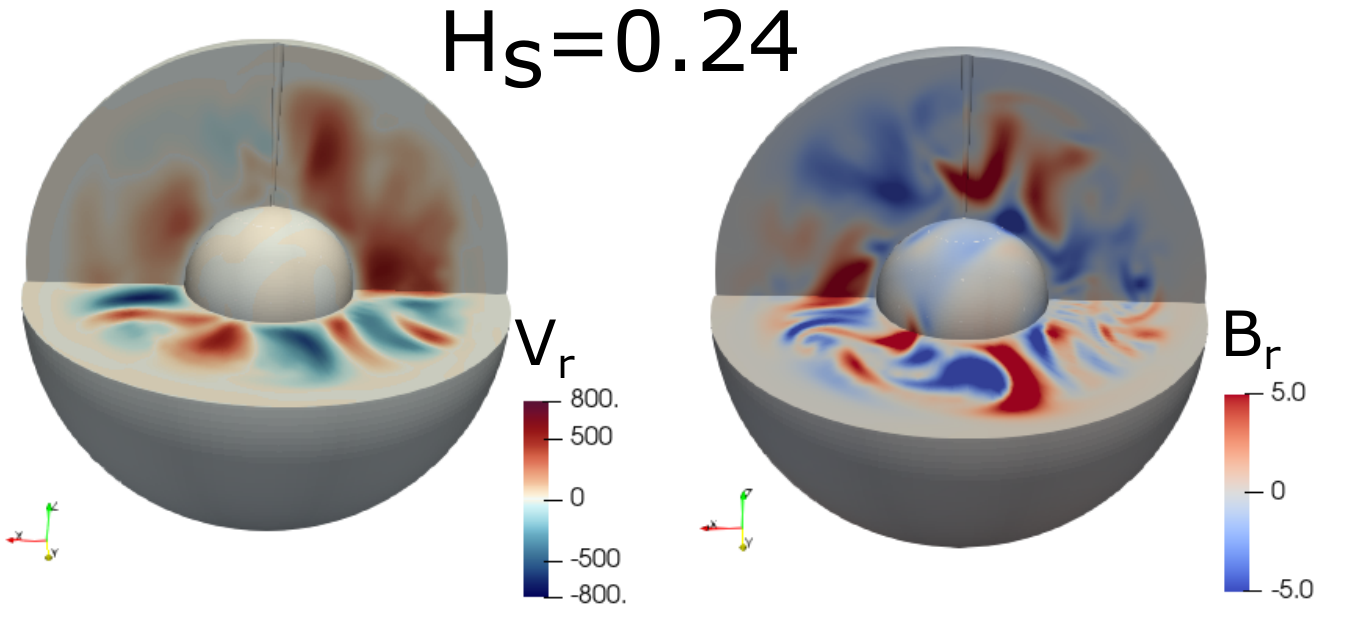}
    \begin{picture}(0,0)
      \put(-200,80){\small (c)} 
    \end{picture}
  \caption{(a) Time averaged dipolar strength $ \fdip$ as a function of the size of the stably-stratified layer $\Hs$ for $\Pm=10$ and $\Ek=10^{-3}$ for different $\Ra$. As the size of the layer $\Hs$ increases, the dipolar component of the magnetic field at the CMB increases due to the skin effect. Error bars correspond to the standard deviation. (Right panels) Visualisation for $\Ra=15 \Rac$ of the radial (left) velocity and (right) magnetic fields for (b) a fully convective system $\Hs=0$, and (c) $\Hs = 0.24 L$. }
  \label{fig:skin_effect}
\end{figure}
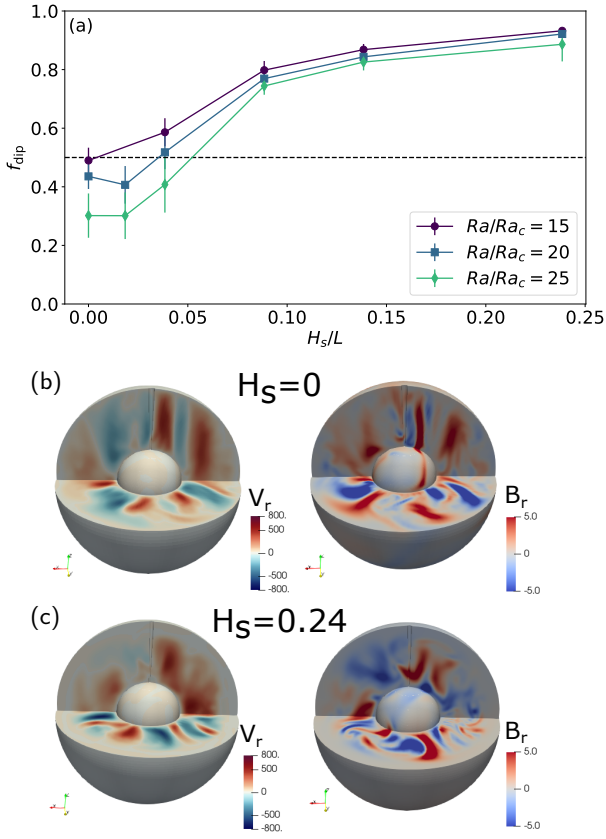

We solve this system of equations using the PARODY-JA code \citep{Dormy1998,Aubert2008}, that uses a pseudo-spectral method in the poloidal and azimuthal directions, and a second-order finite-differences scheme for the radial direction, with a refined number of collocation points near the boundaries. 
The code implements a MPI parallelisation in the radial direction, and it is coupled to the open-source library SHTns \citep{Schaeffer2013} that implements an openMP parallelisation in the angular directions to improve the performance of the spherical harmonic transforms. The resolution in the angular directions is defined such that the ratio between the maximum and minimum values of the magnetic energy spectrum is larger than $50$. The resolution in the radial direction is chosen so that there are at least $10$ points in the Ekman layer at the boundaries.

We introduce an SSL in our system of equations in an \textit{ad-hoc} manner by implementing a static temperature gradient background given by a piecewise function. In the convective region, this gradient is given by the solution of $\laplacian T_s = 0$ for $r_i<r<r_s$, while we propose a simple linear profile for the stably-stratified region $r_s<r<r_o$, leading to \citep{Nakagawa2011,Nakagawa2015,Couston2017,Christensen2018,Gastine2020}
\begin{equation}
    \frac{dT_s}{dr} =
    \begin{cases} 
        -\frac{r_o^2}{r^2} \quad & \text{for} \quad r_i<r<r_s \\
        \Gamma \frac{r-r_s}{\Hs} - \frac{r_o^2}{r_s^2} \frac{r_o-r}{\Hs} \quad & \text{for} \quad r_s<r<r_o
    \end{cases}
    \label{eq:static_temperature_gradient}
\end{equation}
with $\Hs = r_o - r_s$ the size of the SSL, and $\Gamma$ an extra free parameter that controls the maximum stratification strength as $\Nmax/\Omega = \sqrt{\Ra \Ek \Gamma}$. We choose this value so that the maximum stratification strength is between $5$ and $15$, corresponding to a relatively strong stratification. We vary the size of the layer from $\Hs=0$ to $\Hs=0.24 L$. The shape of the static temperature gradient is shown schematically in Fig.~\ref{fig:temperature_profile}. Other choices of the static temperature gradient are possible, such as a continuous profile given by an hyperbolic tangent \citep{Nakagawa2011,Nakagawa2015,Takahashi2019}. The drawback of these choices is that they introduce an extra parameter that determines the size of the transition between the convective and stably-stratified regions. 
This layer can also be modelled using double-diffusive convection, considering both thermal and compositional buoyancies, that leads to different regimes such as classical or finger convection \citep{Monville2019,Guervilly2022}.
In this work, we restrict ourselves to a linear profile for the SSL below the CMB.

On top of the SSL, we explore the effects of an heterogeneous heat flux at the CMB. For this purpose, we impose $q_\CMB(\theta,\phi) = q_o + F(\theta,\phi)$ with $F(\theta,\phi)$ a function proportional to the spherical harmonics $\Ylm$ of degree $l$ and order $m$. In particular, we study the effects of the axisymmetric patterns $Y_1^0$ or $Y_3^0$, and a linear combination of them, such that $F(\theta,\phi) = F(\theta)$. We control the amplitude of this heat flux as 
\begin{equation}
  \dq = \frac{q_{\mathrm{max}} - q_{\mathrm{min}}}{2q_o}.
  \label{eq:heat_flux}
\end{equation}

\section{Outputs}

\begin{figure}
  \centering
  \includegraphics[width=\linewidth]{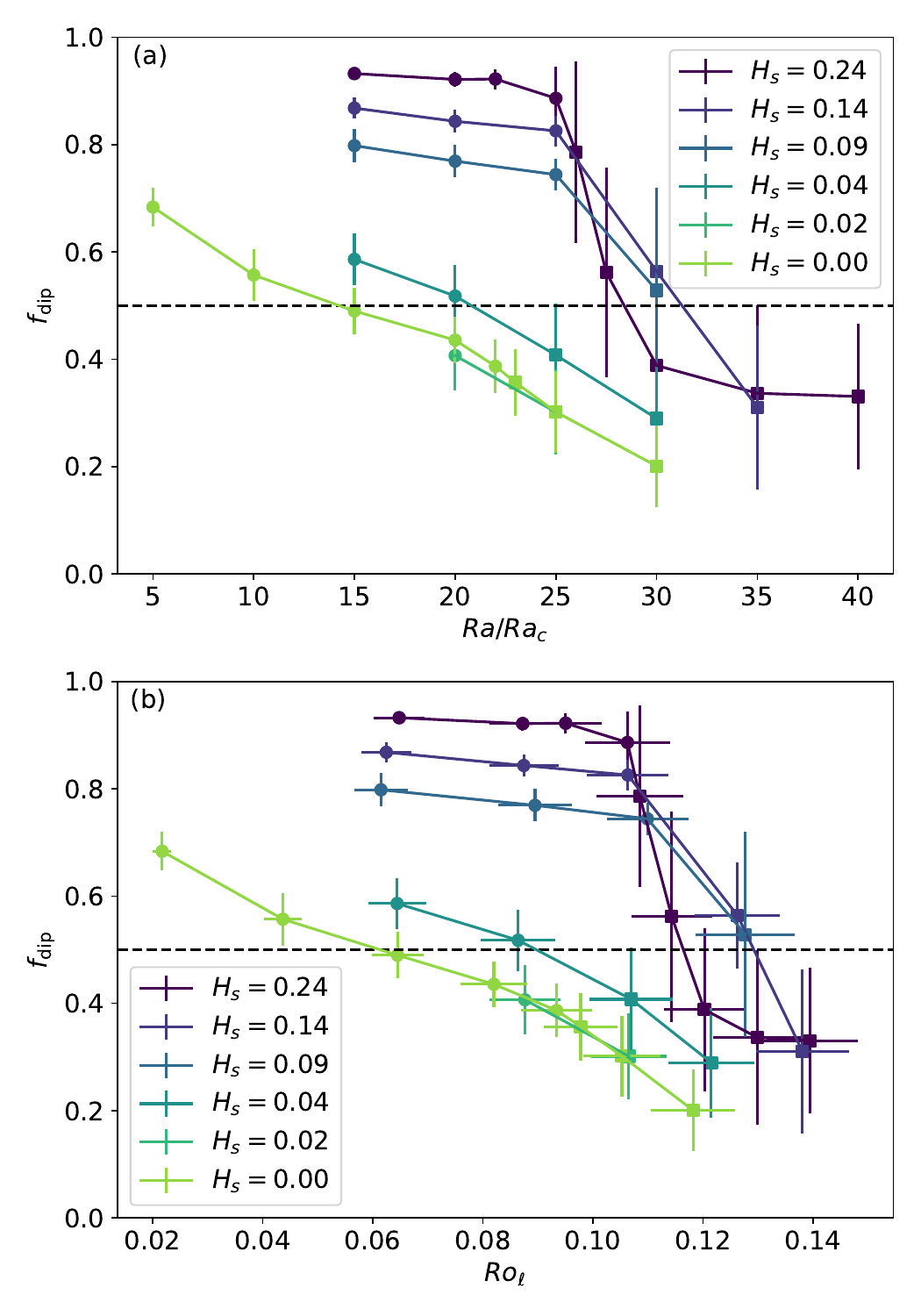}
  \caption[]{Dipolar-multipolar transition for different sizes $\Hs$ of the stably-stratified layer as a function of (a) the Rayleigh number and (b) the local Rossby number for $\Ek=10^{-3}$, $\Pr=1$, and $\Pm=10$. Circles indicate stable solutions, while square markers indicate reversing solutions.
    }
    \label{fig:fdip_ra}
\end{figure}

The code PARODY-JA solves the incompressible MHD Eqs.~\eqref{eq:ns}-\eqref{eq:incompressible} using the poloidal-toroidal decomposition, where the total magnetic field $\Bvec$ is expressed in terms of two scalar potentials $\Bpol$ and $\Btor$ such that $\Bvec = \gradient \times \gradient \times (\Bpol \rvec) + \gradient \times (\Btor \rvec)$. The total magnetic field $\Bvec$ can be described as a linear combination of spherical harmonics $\Ylm$. 
In particular, we can extract the axisymmetric dipole and quadrupole components of the magnetic field at the CMB from the poloidal and toroidal decomposition as 
\begin{align}
\begin{split}
  D &= \Bpol^{\CMB}(l=1,m=0), \\
  Q &= \Bpol^{\CMB}(l=2,m=0).
\end{split}
\label{eq:DQ}
\end{align}
Note that the toroidal component of the magnetic field is zero at the CMB due to the insulating mantle.

The code is expressed in units of the viscous time $t_\nu = L^2/\nu$, which is related to the magnetic diffusion time $t_\eta = L^2/\eta = \Pm t_\nu$. To study magnetic field reversals, it is more appropriate to express the time in terms of the dipole diffusion time 
\begin{equation}
  \tdip = \frac{r_o^2}{\pi^2 \eta} = \frac{\Pm}{\pi^2 (1-\chi)^2} t_\nu,
  \label{eq:tdip}
\end{equation}
with $\chi=0.35$. The dipole diffusion time is typically $\tdip \sim 38.000$ years for Earth's core.

We define the total dimensionless kinetic and magnetic energies as
\begin{gather}
  \Ekin = \frac{1}{2V} \int_V \uvec^2 \mathrm{d}V, \\
  \Emag = \frac{1}{2V} \int_V \left(\frac{\Bvec}{\sqrt{\Ek \Pm}}\right)^2 \mathrm{d}V,
\end{gather}
with $V$ the volume of the spherical shell. The amplitudes of the velocity and magnetic fields are quantified by the Reynolds number $\Rey = UL/\nu$ and the Elsasser number $\Lambda = B^2 / (\rho \mu \eta \Omega)$ with $U$ and $B$ the volume-averaged RMS amplitudes of the velocity and magnetic fields, respectively. For the adimensionalisation used in our simulations, these numbers are computed as 
\begin{gather}
  \Rey = \sqrt{2\Ekin}, \\
  \Lambda = 2 \Pm \Ek \Emag,
\end{gather}
from which we can compute the magnetic Reynolds number as
\begin{equation}
  \Rm = \Pm \Rey.
\end{equation}

To characterize the dipolar-multipolar transition, we define the relative dipole field strength at the outer boundary as 
\begin{equation}
  \fdip = \frac{\sqrt{\Emag^\CMB(l=1)}}{\sqrt{\sum_{l=1}^{12} \Emag^\CMB(l)}}
  \label{eq:fdip}
\end{equation}
with $\Emag^\CMB(l) = \sum_m \Emag^\CMB(l,m)$ the magnetic energy spectrum evaluated at the CMB. The denominator is truncated at $l=12$ to compare with geomagnetic observations, that lead to a typical value for the Earth $\fdip = 0.68$ \citep{Christensen2006,Menu2020}. The transition is typically controlled by the local Rossby number
\begin{equation}
  \Rol = \frac{\lconv \Ro}{\pi},
  \label{eq:Rol}
\end{equation}
with $\lconv = \sum_l l \Ekin(l) / \sum_l \Ekin(l)$ the characteristic spherical harmonic degree of the flow, and $\Ro = \Rey \Ek$ the Rossby number with $\Rey = \sqrt{2 \Ekin}$ the Reynolds number. 

To quantify the equatorial symmetry of the magnetic field, we use the mean hemisphericity at the CMB
\begin{equation}
  \Hcmb = \frac{\mean{B_r^N}^2 - \mean{B_r^S}^2}{\mean{B_r^N}^2 + \mean{B_r^S}^2},
  \label{eq:hemisphericity}
\end{equation}
with $B_r^N$ and $B_r^S$ the radial magnetic field evaluated at the CMB in the northern and southern hemispheres, respectively. For an equatorially symmetric system, the mean hemisphericity is zero, while if the magnetic field is concentrated in one hemisphere, it is $\pm 1$. 
We quantify the symmetry breaking for the flow with respect to the equator from the antisymmetric toroidal and poloidal components of the kinetic energy as 
\begin{equation}
E_{\mathrm{kin}}^{\mathrm{asym}} = \sum_{l+m = 2n} E_{\mathrm{kin}}^{\mathrm{tor}} (l,m) + \sum_{l+m = 2n+1} E_{\mathrm{kin}}^{\mathrm{pol}} (l,m).
\label{eq:Ekasym}
\end{equation}

\section{Results}
\label{sec:results}

\subsection{Parametric study}
\label{subsec:parametric}

We perform a parametric study of the geodynamo by fixing $\Pr=1$, $\Ek = 10^{-3}$, $\Pm=10$, the aspect ratio of the system $\chi = 0.35$, and varying $\Ra$, and the size of the SSL $\Hs = r_o - r_s$ keeping the stratification strength between $\Nmax/\Omega \in (5,15)$. Some simulations are initialised with a fluid at rest, imposing a small temperature and magnetic perturbations, while others are restarted from runs with similar parameters. The relative large values of $\Ek$ and $\Pm$ are chosen to enable an extensive parametric study of the system, allowing us to identify different regimes of geodynamo behaviour. The input and ouptut quantities of the simulations are shown in Table \ref{tab:simulations}.

\begin{table*}[h]
    \centering
    \caption{Table of simulations homogeneous boundary conditions. We report the size of the stably-stratified layer $\Hs$, the normalised Rayleigh number $\Ra/\Rac$, the kinetic energy $\Ekin$, the magnetic energy $\Emag$, the ratio of kinetic and magnetic energies $\Ekin/\Emag$, the magnetic Reynolds number $\Rm$, the antisymmetric kinetic energy $E_k^{\mathrm{asym}}$, the dipolar strength $\fdip$, whether there are reversals, and the final time of the simulations $t_{\mathrm{run}}$ expressed in diffusive time units. All simulations correspond to $\Ek=10^{-3}$, $\Pr=1$, $\Pm=10$, and $r_i/r_o = 0.35$. }
    \label{tab:simulations}
    
    \begin{tabular}{*{12}{c}}
        \toprule
        $\Hs/L$ & $\Ra/\Rac$ & $\Lambda$ & $E_\mathrm{kin}$ & $E_\mathrm{mag}$ & 
        $E_{\mathrm{kin}} / E_{\mathrm{mag}}$ & $\Rm$ & $\Rac$ & $E_k^{\mathrm{asym}}$ & $\fdip$ & reversals? & $t_{\mathrm{run}}$ \\
        \midrule
        0    & 5  & 16.36 &  94  & 818 & 0.12 & 137 & 47.8  & 0.16 & 0.68 & no  & 4.2 \\
        0    & 10 & 24.12 & 247  &1206 & 0.20 & 222 & 47.8  & 0.19 & 0.56 & no  & 2 \\
        0    & 15 & 23.98 & 428  &1198 & 0.36 & 292 & 47.8  & 0.21 & 0.49 & no  & 2 \\
        0    & 20 & 23.04 & 625  &1152 & 0.54 & 353 & 47.8  & 0.22 & 0.44 & no  & 2 \\
        0    & 22 & 17.08 & 759  & 853 & 0.89 & 389 & 47.8  & 0.19 & 0.39 & no  & 5 \\
        0    & 23 & 15.29 & 819  & 764 & 1.07 & 404 & 47.8  & 0.18 & 0.36 & yes & 3.75 \\
        0    & 25 & 12.89 & 929  & 644 & 1.44 & 431 & 47.8  & 0.18 & 0.30 & yes & 8 \\
        0    & 30 & 12.63 &1164  & 631 & 1.84 & 482 & 47.8  & 0.19 & 0.20 & yes & 2 \\
        \hline
        0.02 & 20 & 16.90 & 680  & 845 & 0.81 & 369 & 47.7  & 0.17 & 0.41 & no  & 3 \\
        0.02 & 25 & 11.59 & 962  & 579 & 1.66 & 438 & 47.7  & 0.17 & 0.30 & yes & 6 \\
        \hline
        0.04 & 15 & 22.09 & 417  &1104 & 0.38 & 289 & 48.0  & 0.17 & 0.59 & no  & 4 \\
        0.04 & 20 & 17.25 & 677  & 862 & 0.79 & 368 & 48.0  & 0.16 & 0.52 & no  & 3 \\  
        0.04 & 25 & 11.59 & 995  & 579 & 1.72 & 446 & 48.0  & 0.15 & 0.41 & yes & 9 \\
        0.04 & 30 & 11.96 &1228  & 598 & 2.05 & 495 & 48.0  & 0.16 & 0.37 & yes & 3 \\
        \hline
        0.09 & 15 & 22.48 & 373  &1124 & 0.33 & 273 & 50.22 & 0.18 & 0.80 & no  & 4 \\
        0.09 & 20 & 16.72 & 599  & 836 & 0.72 & 346 & 50.22 & 0.16 & 0.77 & no  & 2.5 \\
        0.09 & 25 & 14.05 & 841  & 702 & 1.20 & 410 & 50.22 & 0.16 & 0.74 & no  & 3 \\
        0.09 & 30 & 9.68  &1175  & 484 & 2.43 & 484 & 50.22 & 0.16 & 0.53 & yes & 6.75 \\
        \hline
        0.14 & 15 & 21.35 & 371  &1067 & 0.35 & 272 & 56.81 & 0.18 & 0.87 & no  & 2.49 \\
        0.14 & 20 & 17.29 & 596  & 864 & 0.69 & 345 & 56.81 & 0.17 & 0.84 & no  & 2.54 \\
        0.14 & 25 & 15.21 & 841  & 760 & 1.11 & 410 & 56.81 & 0.16 & 0.83 & no  & 2.52 \\
        0.14 & 30 & 7.76  &1257  & 387 & 3.24 & 501 & 56.81 & 0.14 & 0.56 & yes & 2.51 \\
        0.14 & 35 & 8.74  &1435  & 437 & 3.28 & 535 & 56.81 & 0.17 & 0.31 & yes & 1.67 \\
        \hline
        0.24 & 15 & 17.43 & 374  & 871 & 0.43 & 273 & 66.4  & 0.18 & 0.93 & no  & 3 \\
        0.24 & 20 & 14.28 & 584  & 713 & 0.82 & 341 & 66.4  & 0.17 & 0.92 & no  & 1.27 \\
        0.24 & 22 & 13.23 & 681  & 661 & 1.03 & 369 & 66.4  & 0.17 & 0.92 & no  & 3 \\
        0.24 & 25 & 9.25  & 927  & 462 & 2.01 & 430 & 66.4  & 0.15 & 0.89 & no  & 7.56 \\
        0.24 & 26 & 9.35  & 979  & 467 & 2.10 & 442 & 66.4  & 0.15 & 0.79 & yes & 5.28 \\
        0.24 &27.5& 7.32  &1100  & 366 & 3.00 & 469 & 66.4  & 0.15 & 0.56 & yes & 3.56 \\
        0.24 & 30 & 7.47  &1212  & 373 & 3.25 & 492 & 66.4  & 0.16 & 0.39 & yes & 3 \\
        0.24 & 35 & 9.75  &1342  & 487 & 2.75 & 518 & 66.4  & 0.19 & 0.34 & yes & 2.25 \\
        0.24 & 40 & 10.72 &1508  & 536 & 2.81 & 549 & 66.4  & 0.20 & 0.33 & yes & 1.5 \\
        \bottomrule
    \end{tabular}
\end{table*}

\begin{table*}[t]
\centering
\caption{Table of simulations with heterogeneous boundary conditions. We report the same parameters as in Table~\ref{tab:simulations}. The heat flux pattern at the CMB is determined by the spherical harmonic $\Ylm$, and the amplitude of the heterogeneity is given by $\dq$ defined in Eq.~\eqref{eq:heat_flux}. All simulations correspond to $\Ek=10^{-3}$, $\Pr=1$, $\Pm=10$, $\Ra=25 \Rac$, and a SSL of size $\Hs=0.24 L$ and $\Nmax / \Omega \approx 10$.}
\label{tab:dynamo_summary}
\begin{tabular}{c c c c c c c c c c c}
\hline
$\Ylm$ & $\dq$ (\%) & $\Lambda$ & $E_{\mathrm{kin}}$ & $E_{\mathrm{mag}}$ & $E_{\mathrm{kin}}/E_{\mathrm{mag}}$ & $\Rm$ & $E_{\mathrm{kin}}^{\mathrm{asym}}$ & $\fdip$ & Reversals? & $t_{\mathrm{run}}$ \\
\hline
$Y_1^0$ & -3.5 & 10.61 & 888 & 530 & 1.67 & 420 & 0.17 & 0.76 & no & 3.10 \\
$Y_1^0$ & 0.0  & 9.59  & 916 & 480 & 1.91 & 427 & 0.15 & 0.90 & no & 7.56 \\
$Y_1^0$ & 0.17  & 9.36  & 932 & 468 & 1.99 & 431 & 0.14 & 0.82 & no & 3.25 \\
$Y_1^0$ & 0.35  & 7.77  & 985 & 389 & 2.53 & 443 & 0.13 & 0.81 & no & 4.45 \\
$Y_1^0$ & 1.05  & 9.12  & 939 & 456 & 2.06 & 432 & 0.14 & 0.82 & no & 3.15 \\
$Y_1^0$ & 1.75  & 10.71 & 887 & 536 & 1.66 & 420 & 0.16 & 0.86 & no & 2.00 \\
$Y_1^0$ & 2.1  & 11.18 & 876 & 559 & 1.57 & 418 & 0.17 & 0.81 & no & 4.00 \\
$Y_1^0$ & 3.5  & 12.23 & 832 & 612 & 1.36 & 407 & 0.19 & 0.78 & no & 6.00 \\
$Y_1^0$ & 8.75  & 12.85 & 828 & 643 & 1.29 & 406 & 0.24 & 0.56 & no & 1.50 \\
\hline
$Y_3^0$ & -5.3 & 11.20 & 880 & 560 & 1.57 & 418 & 0.17 & 0.86 & yes & 9.06 \\
$Y_3^0$ & 0.0  & 9.42  & 922 & 471 & 1.96 & 428 & 0.15 & 0.89 & no  & 7.56 \\
$Y_3^0$ & 0.53  & 10.33 & 884 & 516 & 1.71 & 420 & 0.16 & 0.92 & no  & 5.45 \\
$Y_3^0$ & 2.65  & 10.72 & 867 & 536  & 1.64 & 415 & 0.17 & 0.91 & no  & 13.44 \\
$Y_3^0$ & 5.3  & 9.82  & 908 & 491 & 1.85 & 425 & 0.16 & 0.83 & yes & 29.50 \\
\hline
$Y_1^0 + Y_3^0$ & 1.75  & 8.54  & 957 & 427 & 2.24 & 438 & 0.14 & 0.75 & yes & 27.65 \\
\end{tabular}
\end{table*}

The effects of the SSL in the magnetic field morphology are shown in Fig.~\ref{fig:skin_effect}. The dipolar strength of the magnetic field at the CMB $\fdip$ defined in Eq.~\eqref{eq:fdip} increases with the size of the SSL for a fixed value of $\Ra/\Rac$, where $\Rac$ is the numerically determined critical Rayleigh number for the onset of convection. This is a consequence of the skin effect, which attenuates the contribution of the more spatially structured modes of the magnetic field. The visualisation of the radial velocity and magnetic fields in Figs. \ref{fig:skin_effect} (b)-(c) show how they are attenuated in the presence of the SSL.

Stable dipolar dynamos are typically characterised by a dipolar strength $\fdip>0.5$, whereas lower values correspond to multipolar solutions, in which the dipole component is no longer dominant. The transition between these two dynamo solutions remains a major unresolved issue in our understanding of the geodynamo. Figure \ref{fig:fdip_ra} (a) shows the dipolar-multipolar transition as a function of the Rayleigh number. In the case of a fully convective system ($\Hs=0$), the transition takes place at around $\Ra/\Rac=15$. The introduction of the SSL shifts this transition to larger values, around $\Ra/\Rac=27.5$ for $\Hs=0.24 L$, and makes the dipolar-multipolar transition more abrupt. Figure \ref{fig:fdip_ra} (b) shows the same transition as a function of the local Rossby number defined in Eq.~\eqref{eq:Rol}. The critical value of $\Rol$ for the transition depends on the size of the layer. 
\citet{Zaire2022} proposed that the transition between stable and reversing dynamos is controlled by the ratio between kinetic and magnetic energies, that we report in Table \ref{tab:simulations}. Their study in fully convective systems suggests that the transition takes place at around $\Ekin/\Emag \approx 0.7$. In our case, as the size of the stably-stratified layer increases, this transition seems to shift to larger values, being around $2$ for $\Hs = 0.24 L$.

\begin{figure}
  \centering
  \includegraphics[width=\linewidth]{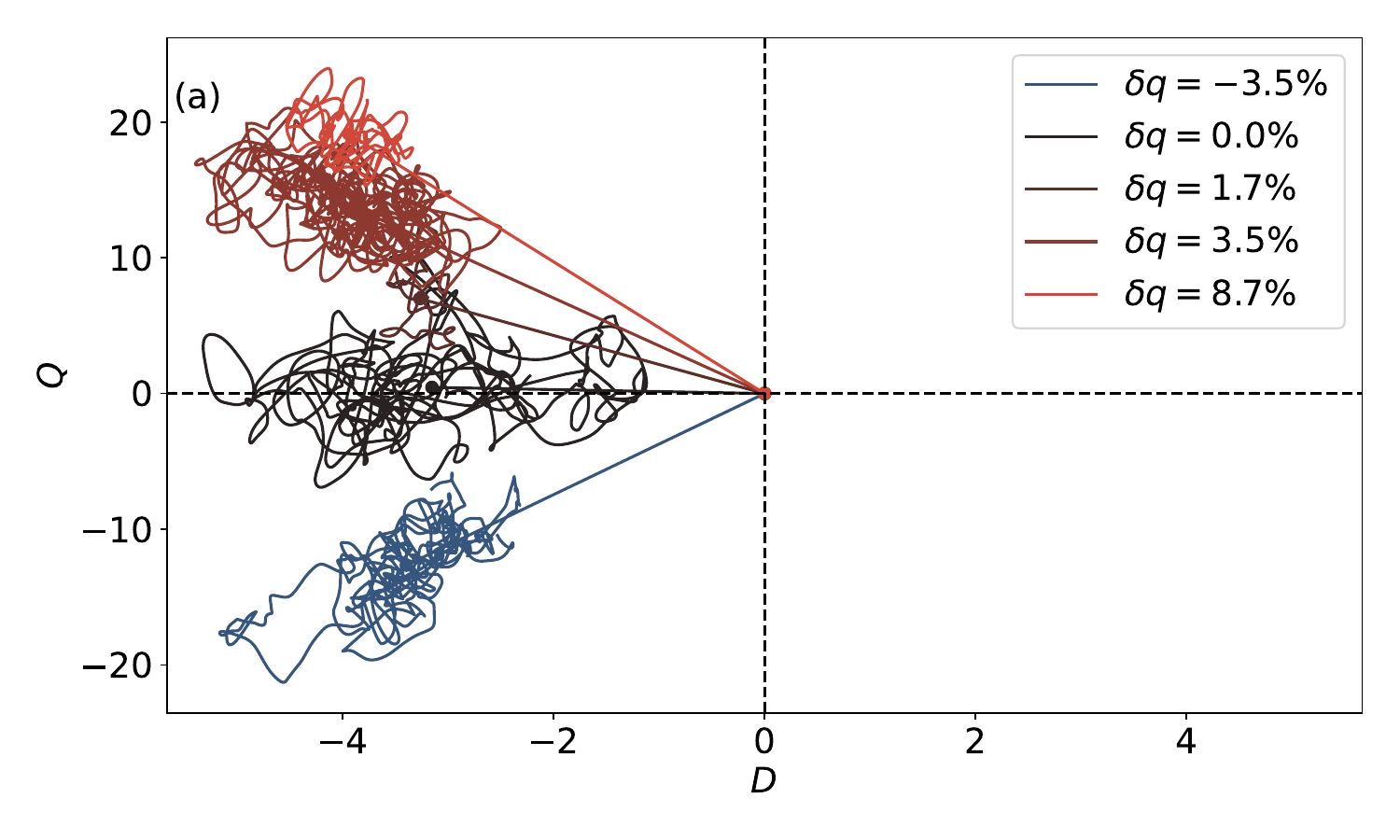}
  \includegraphics[width=\linewidth]{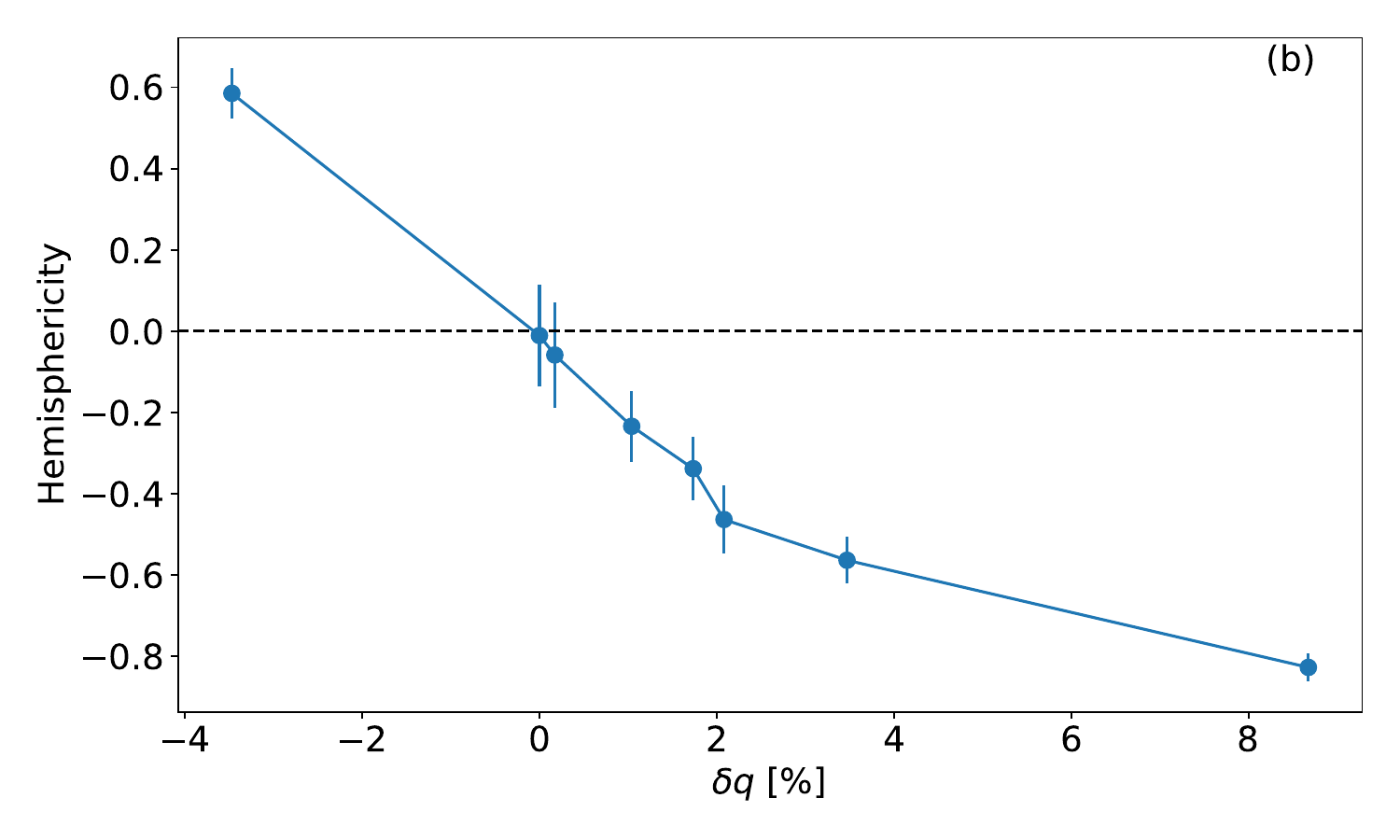}
  \caption[]{(a) Evolution of the magnetic field at the CMB in the D-Q phase space for different amplitudes of the heterogeneous heat flux $\dq$ for the simplest pattern $Y_1^0$. The points correspond to a portion of the time series ($50\tau_\nu$) taken from the statistically steady state of the dynamo solution. (b) Time averaged hemisphericity, defined as in Eq.~\eqref{eq:hemisphericity}, as a function of $\dq$. All simulations correspond to $\Ek=10^{-3}$, $\Pr=1$, $\Pm=10$, and $\Ra / \Rac = 25$. Error bars indicate the standard deviation. 
    }
    \label{fig:hemisphericity}
\end{figure}

\subsection{Effects of an heterogeneous heat flux at the CMB}
\label{subsec:heterogeneous}

Having established the influence of the SSL on the dipolar-multipolar transition, we now turn to exploring the effects of an imposed axisymmetric heterogeneous heat flux at the CMB. We choose the case with a large SSL with $\Hs=0.24 L$ and a value of $\Ra/\Rac=25$ such that it is close to the transition with a stable dipolar solution. The input and output quantities of these simulations are summarised in Table \ref{tab:dynamo_summary}.
We start by imposing the $Y_1^0$ pattern at the CMB, and vary the amplitude of the symmetry breaking $\dq$. 
We verified that when imposing a heat flux at the CMB lower than $10 \%$, the equatorial symmetry of the flow does not change significantly. Figure \ref{fig:hemisphericity} shows the phase diagram of the axial dipole $D$ and quadrupole $Q$ components of the magnetic field at the CMB, defined in Eq.~\eqref{eq:DQ}, for different values of $\dq$ in the presence of an SSL. In the case of no symmetry breaking (black curve), the magnetic field fluctuates around a stable state with a strong dipole component, and a quadrupole that fluctuates around zero. As the amplitude of the heterogeneous heat flux increases (red curve), the quadrupole is enhanced, and the magnetic field fluctuates around a mixed state, and does not display reversals. A movie of the simulation with $\dq = 8.5\%$ can be found in the Supplementary Material \cite{supp}. Moreover, when the amplitude of the heterogeneous heat flux is negative (blue curve), the quadrupole also changes its sign. The straight line connecting to the origin is to indicate the angle between the dipole and quadrupole components.
This enhancement of the quadrupole component is associated with an increase in the hemisphericity of the magnetic field, defined in Eq.~\eqref{eq:hemisphericity} and shown in Fig.~\ref{fig:hemisphericity}-(b). It is remarkable that breaking the heat flux symmetry by only a few percent - leading to a minimal modification of the velocity field ($\Ekin^\mathrm{asym}/\Ekin^\mathrm{tot} < 20$\%) - can result in a strongly hemispherical magnetic field, in line with the results of an $\alpha^2$ model of dynamo \citep{Gallet2009}.

\begin{figure}
  \centering
  \includegraphics[width=\linewidth]{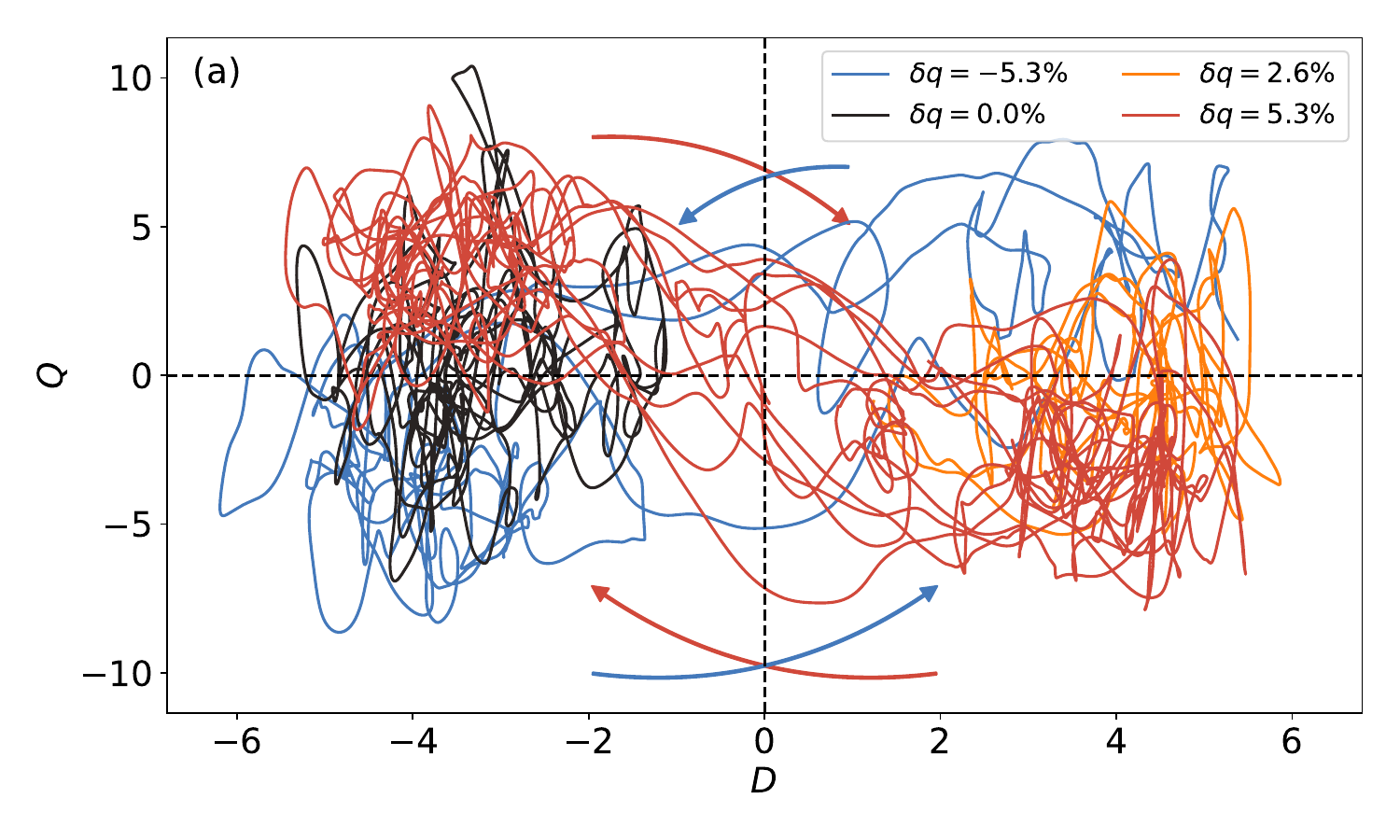}
  \includegraphics[width=\linewidth]{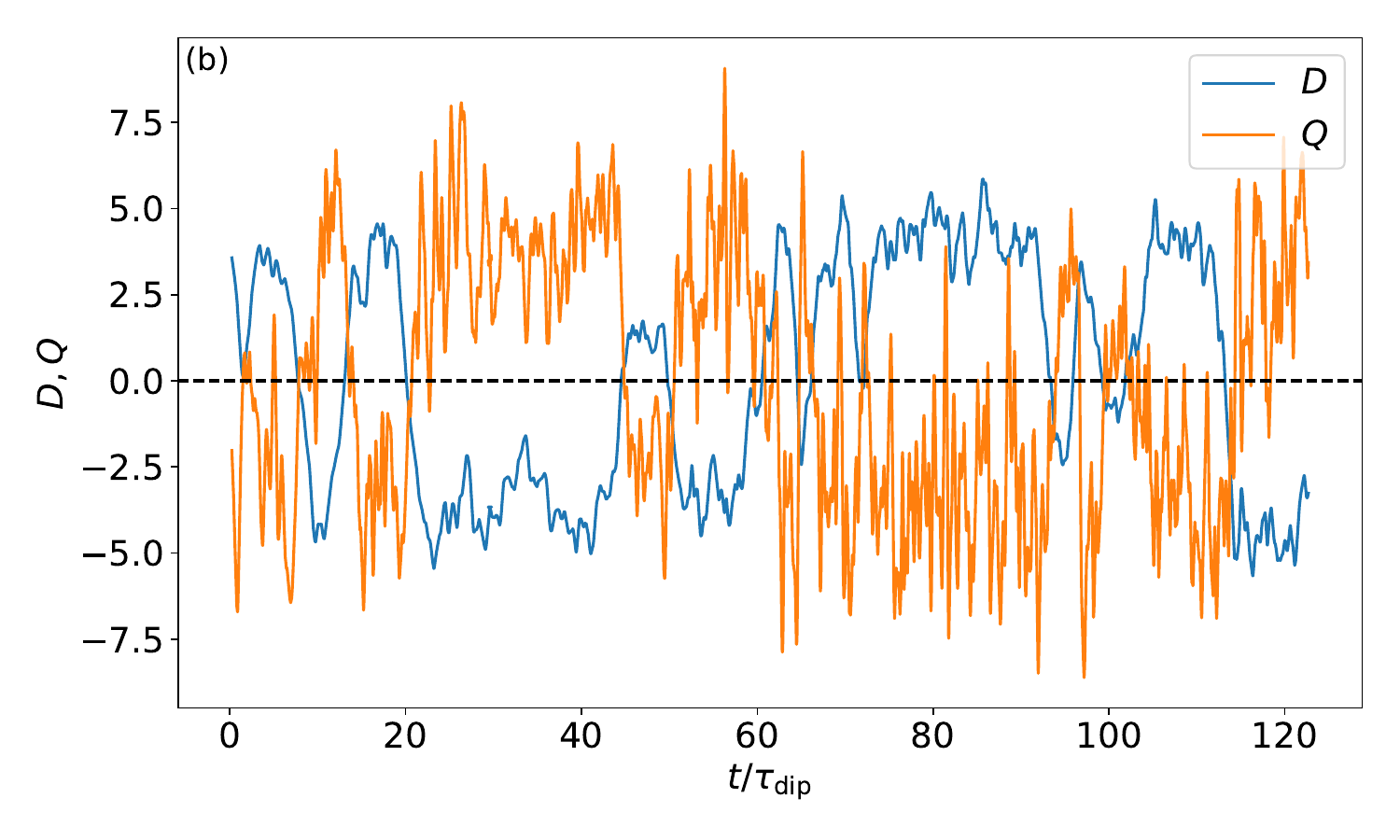}
  \caption[]{
  (a) Evolution of the magnetic field at the CMB in the D-Q phase space for different amplitudes of the heterogeneous heat flux $\dq$ for the pattern $Y_3^0$. Arrows indicate the orientation of the typical trajectories of reversals. (b) Time series of $D$ and $Q$ for the case $\dq = 5.3\%$. All simulations correspond to $\Ek=10^{-3}$, $\Pr=1$, $\Pm=10$, and $\Ra / \Rac = 25$.
    }
    \label{fig:DQ_l3}
\end{figure}

\begin{figure}
  \centering
  \includegraphics[width=\linewidth]{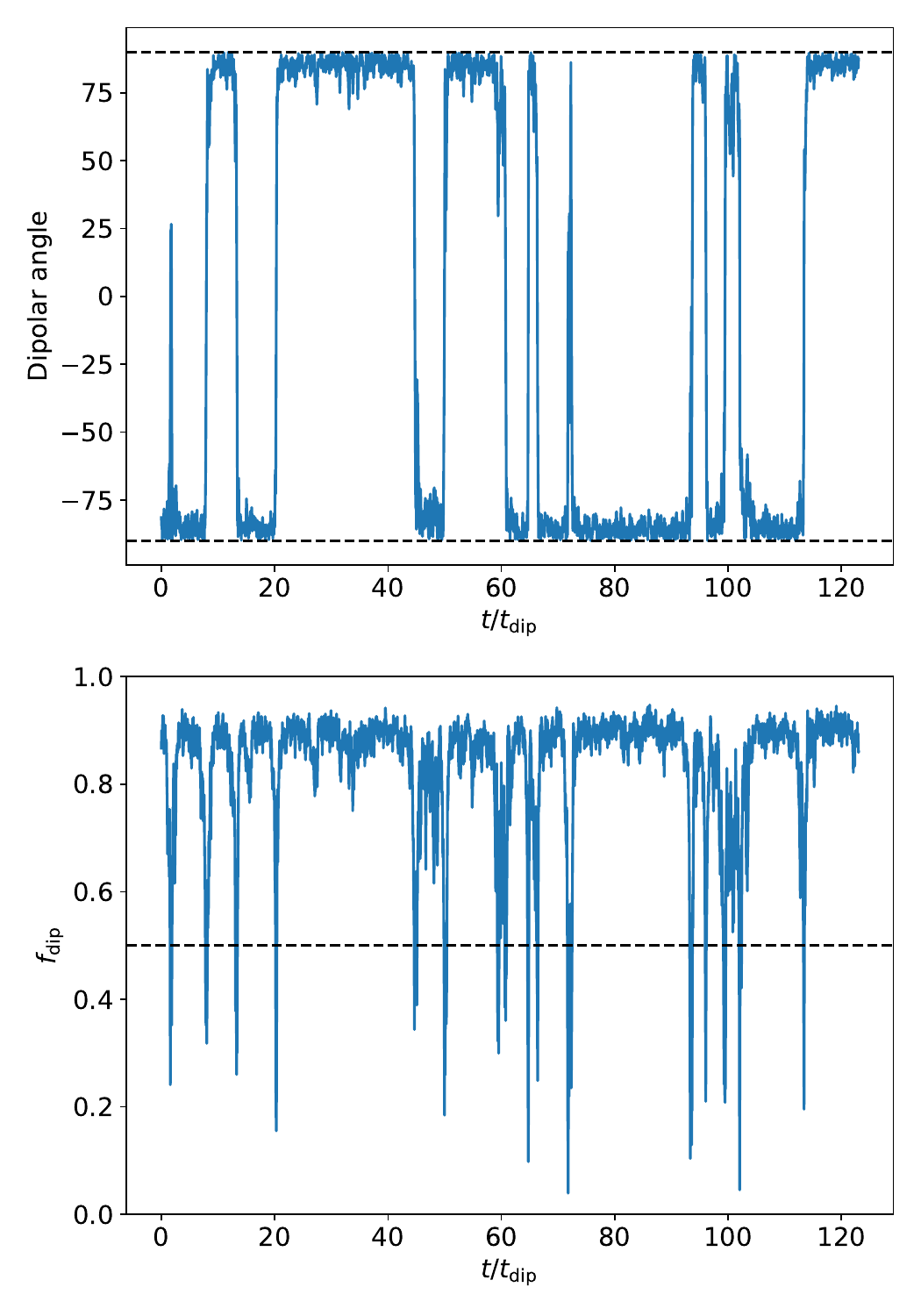}
  \caption[]{Polarity reversals as a function of dipolar diffusion time for a symmetry breaking $Y_3^0$ with an amplitude $\dq=5.3 \%$, $\Ek = 10^{-3}$, $\Pm=10$, $\Pr=1$ and $\Ra = 25 \Ra_c$ in the presence of a stably-stratified layer.  
    }
    \label{fig:reversals}
\end{figure}

We also investigate the effects of a more complex heterogeneous heat flux pattern at the CMB given by the spherical harmonic $Y_{3}^{0}$. 
This choice is motivated primarily by theoretical and methodological considerations: $Y_3^0$ provides a simple axisymmetric pattern with increased latitudinal complexity compared to $Y_1^0$, allowing us to investigate the effects of higher-degree axisymmetric heterogeneity.
The results of the simulations are shown in Fig.~\ref{fig:DQ_l3}. 
As the amplitude of the heterogeneous heat flux increases, the system eventually reaches a regime in which there are magnetic reversals, switching from a $\Bvec \rightarrow -\Bvec$ solution, where both the dipole and quadrupole components change sign (red curve). Arrows indicate the typical trajectories of reversals. 
The time series of the dipole and quadrupole for $\dq = 5.3\%$ is shown in Fig.~\ref{fig:DQ_l3} (b). 
It is found systematically that during the reversal process, the dipole reverses first, and the quadrupole follows slightly later. A movie of this simulation can be found in the Supplementary Material \cite{supp}. Reversals follow the opposite trajectory when the amplitude of the heterogeneous heat flux is negative (blue curve).
A long evolution of the dipolar angle and the dipolar strength $\fdip$ for $\dq = 5.3\%$ is shown in Fig.~\ref{fig:reversals}. In the period of $120$ dipole diffusion times, we observe several non-periodic chaotic reversals of the magnetic field. During the reversal, the dipolar strength decreases from $\fdip \approx 0.9$ to a value $\fdip<0.5$, consistent with the reduction of the dipole component during this event shown in Fig.~\ref{fig:DQ_l3}(b). The high value of $\fdip$ between reversals can be attributed to the presence of the SSL, allowing for a strong dipolar magnetic field that reverses, in contrast with the magnetic reversals observed in the absence of a layer in the multipolar regime, dominated by high order modes. 
We also verified the robustness of these solutions by introducing small variations (up to $10 \%$) in both the Rayleigh number and the amplitude of the heterogeneous heat-flux pattern $Y_3^0$. In these cases, we recover reversals with similar properties to those shown in Figs.~\ref{fig:DQ_l3} and \ref{fig:reversals}.

\begin{figure}
  \centering
  \includegraphics[width=\linewidth]{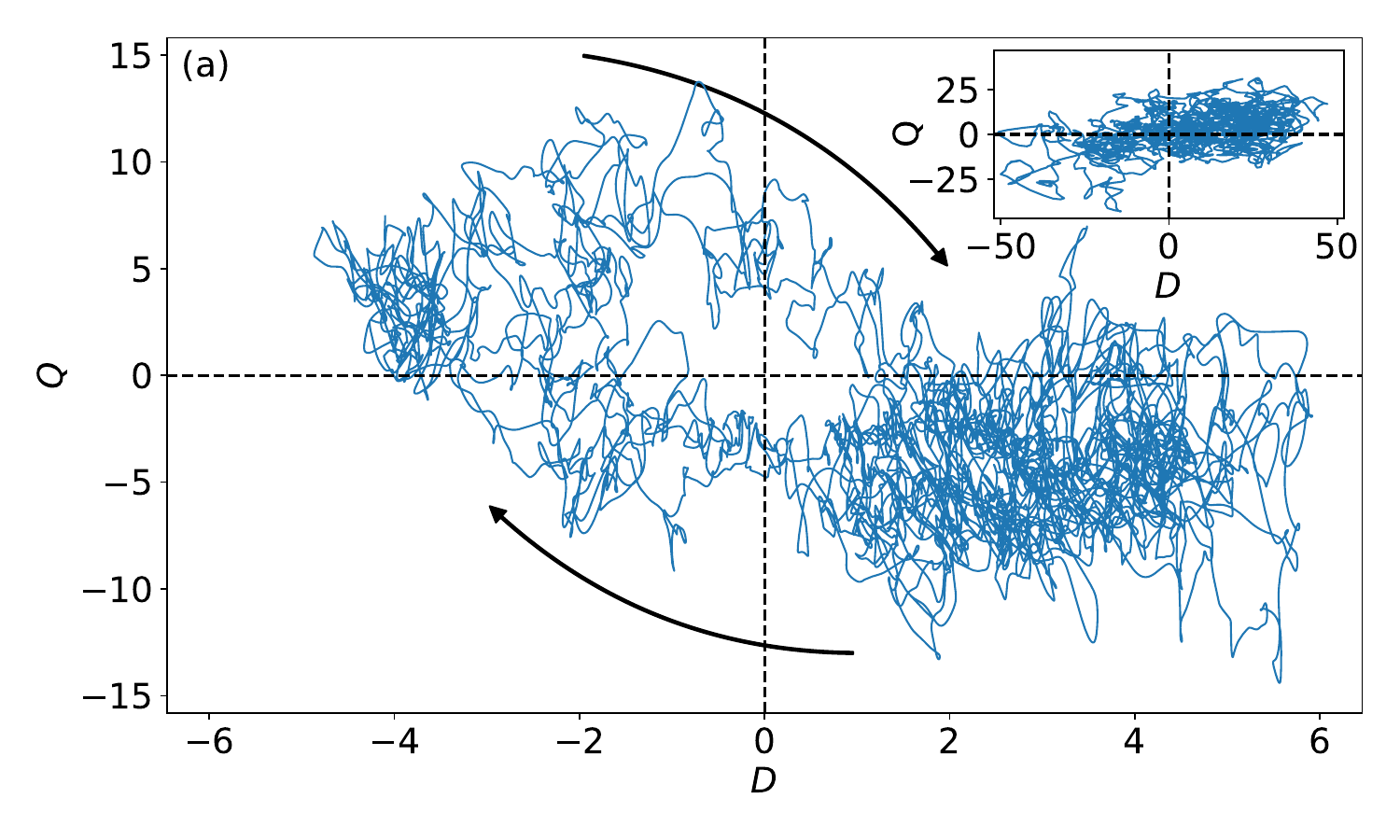}
  \includegraphics[width=\linewidth]{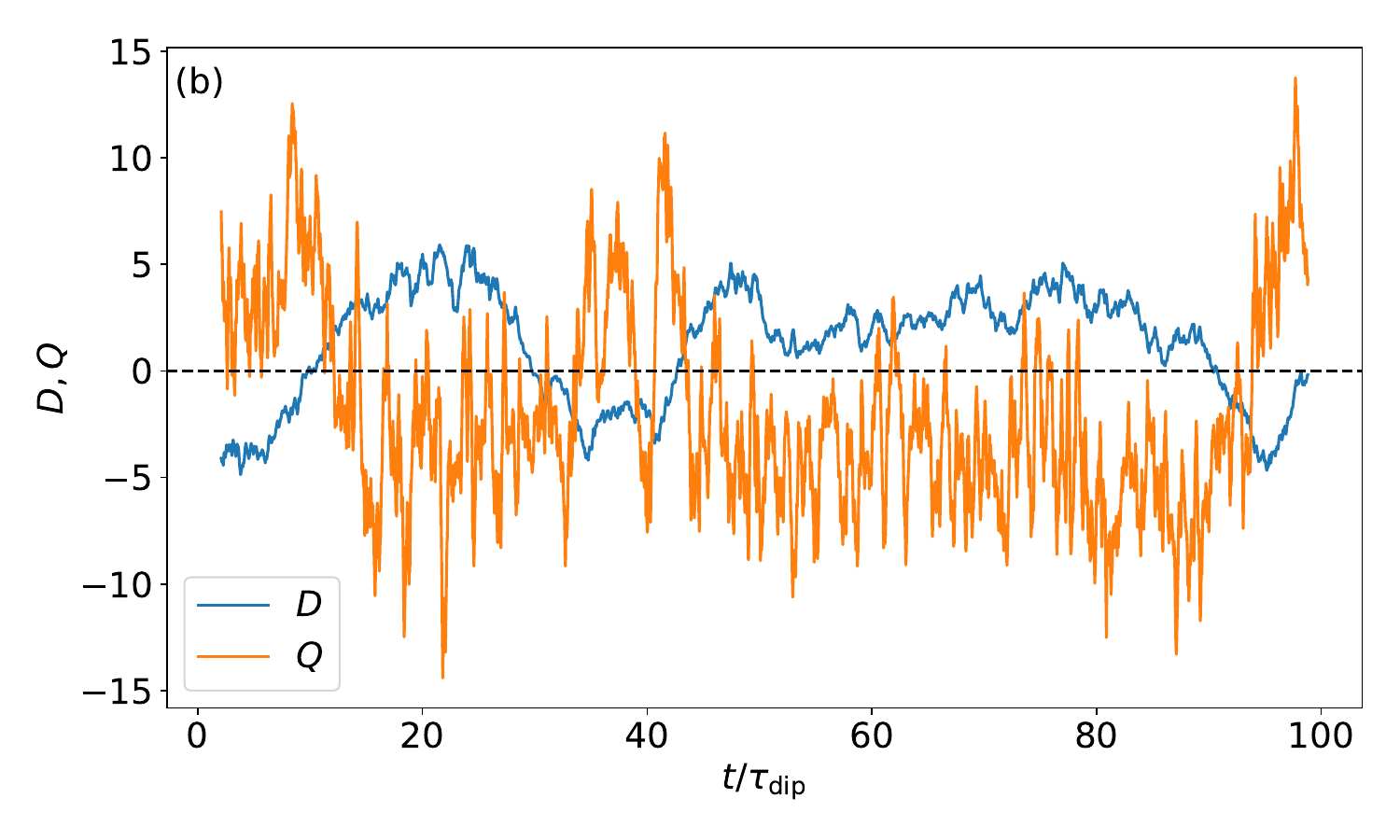}
  \caption[]{
  Evolution of the dipole and quadrupole magnetic field at the CMB for a linear combination of the heterogeneous heat flux patterns $Y_1^0 + Y_3^0$ with an amplitude $\dq = 1.75\%$. The arrows indicate the orientation of reversals. The inset shows a typical simulation with reversals in a fully convective domain. All simulations correspond to $\Ek=10^{-3}$, $\Pr=1$, $\Pm=10$, and $\Ra/\Rac = 25$. 
    }
    \label{fig:DQ_plus}
\end{figure}

We can also consider a linear combination of the two patterns $Y_1^0$ and $Y_3^0$ as a heterogeneous heat flux at the CMB. 
Figure \ref{fig:DQ_plus} shows the evolution of the magnetic field in the D-Q phase space for a linear combination of these two patterns, both with the same amplitude and a total flux variation $\dq=1.75\%$. A movie of this simulation can be found in the Supplementary Material \cite{supp}. 
This heterogeneous heat flux leads to polarity reversals of the magnetic field, with a trajectory that is qualitatively similar to the one described by $Y_3^0$, but here the time delay between the reversal of the dipole and quadrupole is larger, and the trajectories deviate significantly from the origin. 
The arrows in the figure show the direction of the reversal. The phenomenology observed here differs from the one described in the absence of a SSL, where both the dipole and quadrupole components fluctuate around the origin, as shown in the inset of Fig.~\ref{fig:DQ_plus}(a). 
We performed also a set of simulations (not shown here) to explore if it is possible to recover reversals in the absence of a SSL and for $\Ra/\Rac = 5$. For high amplitudes of $\dq$, we do not recover dynamo solutions as a consequence of a high symmetry breaking of the flow. For low amplitudes of $\dq$ we do observe an enhancement of the quadrupolar component of the magnetic field. In the next section, we provide an interpretation of this phenomenology based on a low-dimensional model and the difference of the dipole and quadrupole growth rates.

\section{Discussion}
\label{sec:discussion}

The phenomenology presented in the previous section, involving the emergence of hemispheric dynamos and polarity reversals due to an equatorial symmetry breaking, can be interpreted using low-dimensional models. 
These models consider that the axial dipole $D$ and axial quadrupole $Q$ are the two dominant components of the magnetic field, which interact in a non-linear manner coupled by the hydrodynamic properties of the system. Taking into account the $\Bvec \rightarrow -\Bvec$ invariance of the MHD equations, this model is written as \citep{Petrelis2008,Petrelis2009} 
\begin{align}
\begin{split}
\dot{D} &= \sigma_D D + \alpha Q + C_{11} D^3 + C_{12} D^2 Q + C_{13} D Q^2 + C_{14} Q^3 \\
\dot{Q} &= \sigma_Q Q + \beta D + C_{21} D^3 + C_{22} D^2 Q + C_{23} D Q^2 + C_{24} Q^3 
\label{eq:DQ_model}
\end{split} 
\end{align}
where all the parameters $C_{ij}$, $\alpha$, $\beta$ and $\sigma_{D,Q}$ are real numbers. In particular, the coefficients $\sigma_{D}$ and $\sigma_Q$ determine the dipole and quadrupole growth rates in the limit in which these two modes are decoupled and the non-linear interactions are negligible.
When $\sigma_D$ and $\sigma_Q$ are similar, an equatorial symmetry breaking will result either in periodic polarity reversals or in hemispheric dynamos \citep{Gallet2009}. 
In the full model, stochastic term is introduced to account for the effects of turbulence \citep{Petrelis2008}. 
Alternatively, Eqs~\eqref{eq:DQ_model} can be coupled to a third mode modelling the asymmetric velocity mode generated by the heterogeneous heat flux, resulting in a deterministic model for chaotic inversions \citep{Gissinger2012a}.

\begin{figure}
  \centering
  \includegraphics[width=\linewidth]{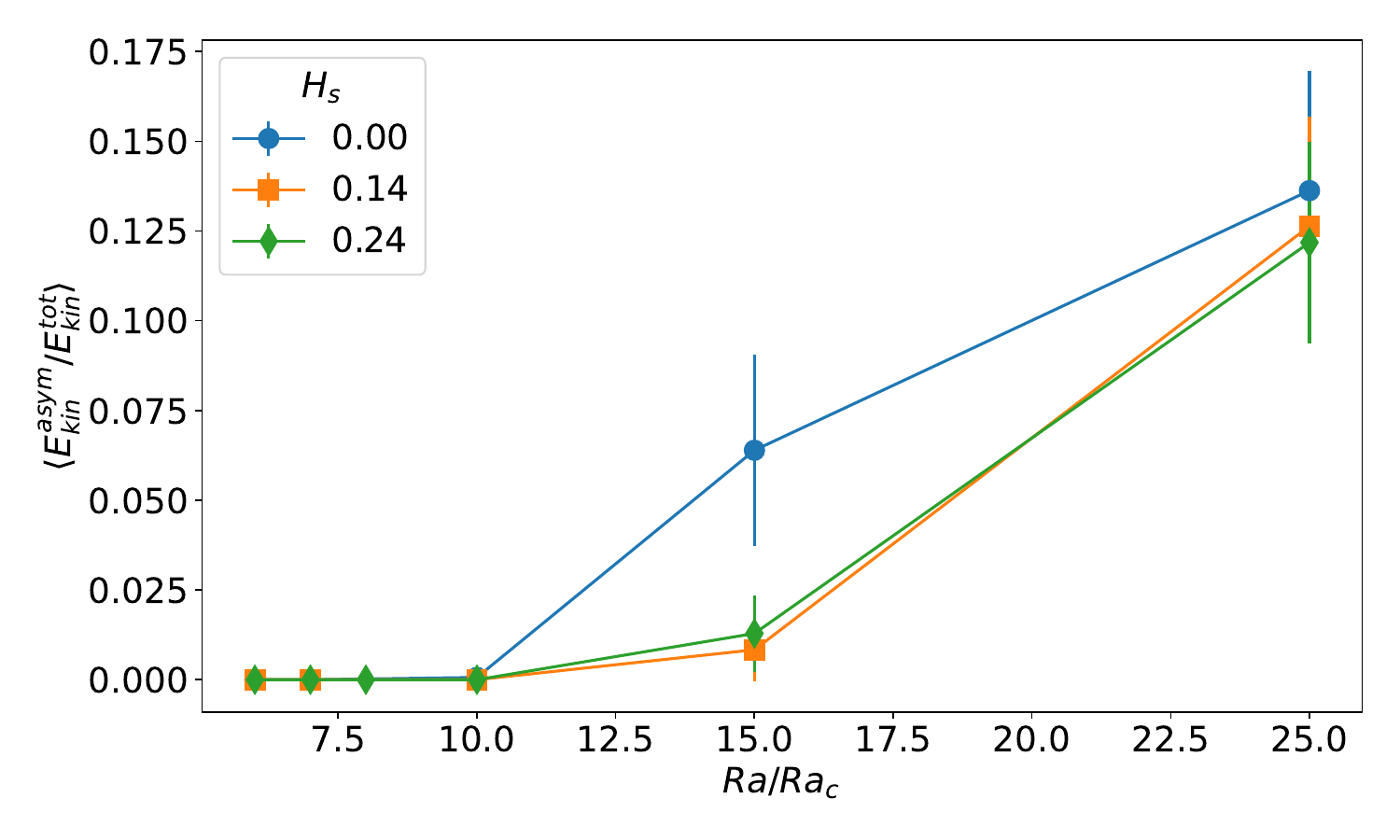}
  \includegraphics[width=\linewidth]{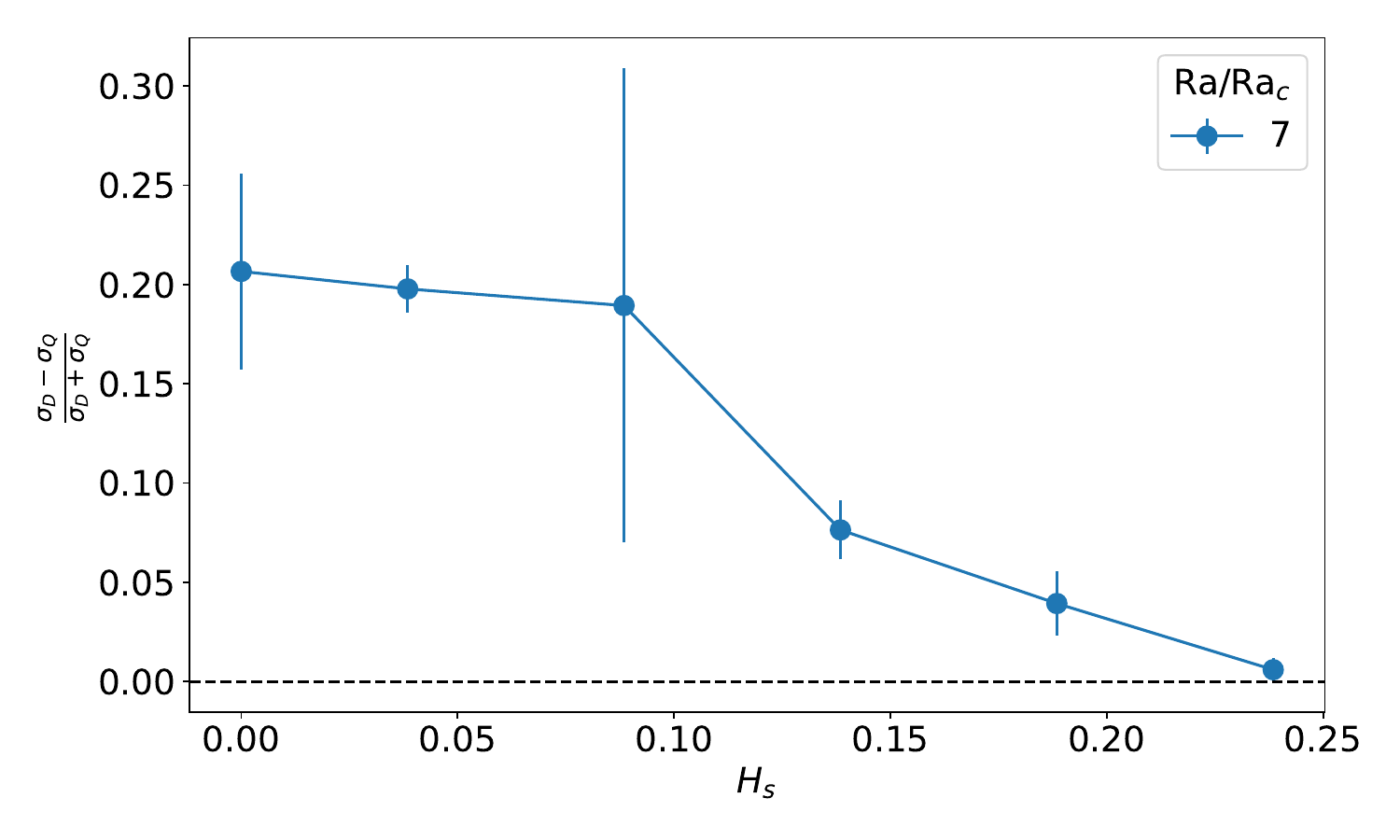}
  \caption{Kinematic dynamo simulations for different sizes of the stably-stratified layer $\Hs$ and different Rayleigh numbers, for fixed $\Ek=10^{-3}$, $\Pr=1$, and $\Pm=10$. Time-averaged (a) relative antisymmetric kinetic energy and (b) relative growth rate difference between the dipole and quadrupole. Error bars represent the standard deviation. 
    }
    \label{fig:kinematic}
\end{figure}

It has been shown that such low-dimensional models based on dipole-quadrupole interactions can reproduce fairly well the magnetic field reversals observed on Earth \citep{Petrelis2009,Petrelis2011}
and in experimental dynamos \citep{Berhanu2007,Gissinger2010a}. However, this behaviour has never been observed, so far, in numerical modelling of the geodynamo, even in the presence of a heterogeneous heat flux \citep{Gissinger2012}. Interestingly, it is therefore the presence of an SSL in the present simulations that gives rise to this long-anticipated phenomenology.
This could be a consequence of the attenuation of fast modes of the magnetic field due to the skin effect in the SSL. 
A more appealing interpretation involves the relative growth rates of the dipole and quadrupole components and their non-linear interaction. 
To understand the behaviour of these quantities, we perform DNS of the kinematic dynamo model, which corresponds to the same system of equations \eqref{eq:ns}-\eqref{eq:incompressible} eliminating the Lorentz force from Eq.~\eqref{eq:ns}. We point out that the velocity field still evolves in time under the same boundary conditions (including the heterogeneous CMB heat flux) as in the self-consistent simulations, but now independently of the magnetic field. The magnetic field therefore responds passively to the temporal fluctuations of this flow.
We first focus on the hydrodynamic properties of the system, in particular on the equatorial symmetry of the flow as a function of the Rayleigh number. 
Figure~\ref{fig:kinematic}(a) shows the relative antisymmetric kinetic energy for different values of the size of the SSL defined in Eq.~\eqref{eq:Ekasym}.
For low values of $\Ra / \Rac$, this quantity is zero indicating that the flow is symmetric with respect to the equator. In this regime, the dipole and quadrupole components are independent as they are decoupled. At around $\Ra=10 \Ra_c$, the antisymmetric kinetic energy undergoes a transition and the equatorial symmetry of the flow is broken. When this happens, $D$ and $Q$ are coupled. When the two modes have similar growth rates, this favours magnetic reversals. 
We compute the dipole $\sigma_D$ and quadrupole $\sigma_Q$ growth rates as
\begin{gather}
D(t) = D_0 \exp(\sigma_D t), \\
Q(t) = Q_0 \exp(\sigma_Q t).
\end{gather}
It is interesting to note that this transition occurs simultaneously with the dipolar-multipolar transition in the fully convective system, and explains why magnetic reversals are obtained in this regime. This result is consistent with \citet{Garcia2017}, where it was shown that the loss of dipolarity is associated to a breaking of the equatorial symmetry of the flow.

More interestingly, we now focus on the effect of the SSL on the kinematic growth rates of the dipole and quadrupole modes below the symmetry breaking threshold, in the regime where the two linear growth rates remain decoupled. 
Figure \ref{fig:kinematic}(b) shows the dipole and quadrupole growth rates for $\Ra=7 \Rac$ for different values of $\Hs$. When the size of the SSL increases, the relative difference between the kinematic growth rates of the dipolar and quadrupolar modes decreases. 
This behaviour can be understood with the prediction by \citet{Proctor1977}, who showed that in $\alpha^2$-dynamos surrounded by a perfectly conducting, motionless outer layer, a degeneracy arises between the growth rates of the dipole and quadrupole components. 
In this context, the SSL beneath the core-mantle boundary can be interpreted as an effective conducting boundary, leading to a similar degeneracy as its thickness increases \citep{Favier2013}. These conditions strongly favour the emergence of the low-dimensional dynamics described above: as the dipole and quadrupole growth rates $\sigma_D$ and $\sigma_Q$ approach one another, even a weak symmetry breaking induces a spectacularly strong coupling between the two modes, enhancing the dynamics predicted by low-dimensional models - including both chaotic reversals and the development of hemispherical magnetic fields.

\section{Conclusions}
\label{sec:conclusions}

In this work, we studied the effects of a stably-stratified layer (SSL) below the core-mantle boundary (CMB) on the morphology of the magnetic field, with particular focus on polarity reversals. Using direct numerical simulations of the incompressible magnetohydrodynamics equations in a rotating spherical shell driven by convection, we showed that the SSL enhances the dipolar strength of the magnetic field at the CMB due to the skin effect, that attenuates rapidly fluctuating fields.
As a result, the transition from dipolar to multipolar dynamos occurs at higher values of the Rayleigh number when the SSL is present, and the transition becomes more abrupt.

By imposing both an SSL and a heterogeneous heat flux at the CMB to break the equatorial symmetry of the flow, we observed a variety of dynamo regimes, including hemispheric dynamos and polarity reversals. 
Hemispheric dynamos arise from an increase in the quadrupolar component of the magnetic field, which is due to the imposed heterogeneous heat flux and the resulting weak equatorial symmetry breaking of the flow.
For the parameters studied in this work, we found that the simplest heat flux pattern, described by the spherical harmonic $Y_1^0$, controls the hemisphericity of the magnetic field. In contrast, a more complex pattern $Y_3^0$ is responsible for triggering polarity reversals. When a linear combination of these two patterns is imposed, reversals follow a clear trajectory in the dipole-quadrupole ($D-Q$) phase space, with the dipole reversing systematically before the quadrupole. This reversal behaviour differs from that observed in fully convective systems within the multipolar regime, where both the dipole and quadrupole components fluctuate without following a well-defined trajectory in the $D-Q$ plane. We also highlight that between the reversals, the dipolar strength of the magnetic field takes relatively large values $\fdip \approx 0.9$, that are typically not observed in simulations.

We interpret these results and the role of the SSL by means of kinematic dynamo simulations. In fully convective dynamos, we associate the dipolar-multipolar transition with a hydrodynamical instability - controlled by the Rayleigh number - that breaks the equatorial symmetry of the flow. When the flow remains symmetric with respect to the equatorial plane, the dipole and the quadrupole become decoupled, and the growth rates of the dipole $\sigma_D$ and the quadrupole $\sigma_Q$ approach each other as the thickness of the SSL increases. This behaviour is consistent with the prediction by \citet{Proctor1977}, who showed that in $\alpha^2$-dynamos surrounded by a perfect conductor, a degeneracy arises between the dipolar and quadrupolar growth rates.

The mechanism based on the interaction of the dipole and quadrupole components of the magnetic field provides an explanation for the mechanism of magnetic reversals observed in the geodynamo. It explains the observations of \citet{Garcia2017}, where it was shown that the loss of dipolarity is associated to a breaking of the equatorial symmetry of the flow. Recent studies of \citet{Aubert2025} and \citet{Jones2025} provide other alternative for which the geodynamo model can reproduce Earth-like magnetic reversals. In particular, \citet{Aubert2025} provide a kinematic mechanism controlled by the relative strength of subsurface upwelling and horizontal circulation at the core surface. In \citet{Jones2025}, the authors find a regime of parameters in which the magnetic field reverses even in the low inertia limit by fixing the ratio between $\Pm$ and $\Pr$. Our results propose a different mechanism based on a low-dimensional model used to explain reversals in dynamo experiments \cite{Petrelis2009,Petrelis2008}. The simulations showing reversing solutions in this work operate close to the dipolar-multipolar transition, but we expect this mechanism to apply also for low-inertia systems.

This interpretation of the SSL, and its critical role in reducing the difference between dipolar and quadrupolar growth rates ($\sigma_D - \sigma_Q$), provides an elegant resolution to a long-standing question regarding the robustness of magnetic field reversals. Indeed, despite considerable variability in core conditions - such as inner-core growth, fluctuations in the heat flux, and other long-term evolutionary processes - the presence of a substantial SSL acts as a stabilizing factor. By maintaining the near-degeneracy of $\sigma_D$ and $\sigma_Q$, it ensures that the low-dimensional dynamics driven by the dipole-quadrupole interaction remains robust, even in the presence of other complex small scale modes. Consequently, even small symmetry breaking effects become sufficient to sustain the observed reversal behaviour over geological timescales, making Earth's dynamo remarkably resilient to changes in its internal properties.
Finally, we remark the results shown in this work were obtained for a relatively high value of the Ekman number of $\Ek=10^{-3}$ and the magnetic Prandtl number $\Pm=10$, which are quite far from realistic values for the Earth of $\Ek=10^{-15}$ and $\Pm=10^{-6}$. Further studies could investigate the dependence of this phenomenology with the Ekman number and also the magnetic Prandtl number. 

\section*{Acknowledgments}
We acknowledge support from Agence Nationale de la
Recherche (Grant No. ANR-19-CE31-0019-01).
This project was provided with computer and storage resources by GENCI at CINES thanks to the grant A0170515642 on the supercomputer Adastra GENOA. 
This work was also granted access to the HPC resources of MesoPSL funded by the Region Île-de-France.

\bibliographystyle{cas-model2-names.bst}
\bibliography{bibliography}

\end{document}